\documentclass[12pt,a4paper]{article}
\usepackage{amsmath,amsfonts,amsthm,graphicx,lscape}
\usepackage[dvips]{psfrag}
\usepackage{cite}

\usepackage[normalem]{ulem}

\usepackage{textcomp}
\usepackage{amsmath,amsthm,amssymb,mathrsfs,cases}
\usepackage{amsfonts,graphicx,lscape}

\newcommand{\vt}[1]{\mbox{\boldmath$#1$}}
\newcommand{\sca}[2]{\langle #1, #2 \rangle}

\usepackage{accents}
\makeatletter
\def\widebar{\accentset{{\cc@style\underline{\mskip10mu}}}}
\makeatother

\numberwithin{equation}{section} 
\newtheorem{theorem}{Theorem}[section]
\newtheorem{proposition}[theorem]{Proposition}

\def\beqa{\begin{eqnarray}}
\def\enqa{\end{eqnarray}}
\def\beq{\begin{equation}}
\def\enq{\end{equation}}

\begin{document}
\title{
Integrable semi-discretizations 
of the Davey--Stewartson system 
and a \mbox{$(2+1)$}-dimensional
Yajima--Oikawa system.\ II
}
\author{Takayuki \textsc{Tsuchida}
}
\maketitle
\begin{abstract} 
%
This is a continuation of our previous paper 
arXiv:1904.07924, 
which is 
devoted to the construction of integrable 
semi-discretizations 
of the Davey--Stewartson system 
and a 
\mbox{$(2+1)$}-dimensional 
Yajima--Oikawa system; 
in 
this series of papers,  
we refer to 
a discretization of one of the two spatial variables 
as a semi-discretization. 
In this paper, we 
construct 
an integrable 
semi-discrete Davey--Stewartson system, 
which is essentially different from 
the semi-discrete Davey--Stewartson system
proposed in the previous paper 
arXiv:1904.07924. 
We first 
obtain 
integrable semi-discretizations of 
the two elementary flows 
that compose 
the Davey--Stewartson system by 
constructing their Lax-pair representations 
and show that 
these two elementary flows 
commute 
as in the continuous case. 
Then, we 
consider a linear combination of the two elementary flows 
to obtain 
a new integrable 
semi-discretization of the Davey--Stewartson system. 
%
Using a 
linear 
transformation of the 
continuous 
independent variables, 
one of the two 
elementary Davey--Stewartson flows 
can be identified 
with 
an integrable 
semi-discretization of the \mbox{$(2+1)$}-dimensional 
Yajima--Oikawa system 
proposed in 
https://link.aps.org/doi/10.1103/PhysRevE.91.062902 .
\end{abstract}
%
%
\newpage
\noindent
\tableofcontents

\newpage
\section{Introduction}
%
As a continuation of 
our previous paper~\cite{Tsuchida19}, 
we 
consider 
the problem of 
how to discretize one of the two spatial variables 
in the Davey--Stewartson system~\cite{DS74}. 
Both the 
Davey--Stewartson system~\cite{DS74} (also
referred to  as the Benney--Roskes system~\cite{Benney69}) 
and the Calogero--Degasperis system~\cite{Calo76}
are 
integrable 
\mbox{$(2+1)$}-dimensional generalizations of 
the nonlinear Schr\"odinger equation~\cite{Zakh}. 
The integrability of 
the Davey--Stewartson system 
was established by 
Ablowitz and Haberman in 1975~\cite{Hab75} 
(also see~\cite{Morris77,Anker,Ab78,Cornille}), 
who provided its 
Lax-pair representation~\cite{Lax}; 
the Lax-pair representation 
in \mbox{$2+1$} dimensions 
is 
also referred to as 
the Manakov triad 
representation~\cite{Manakov_triad}, 
particularly when it is expressed in operator form. 

The Davey--Stewartson system~\cite{DS74}
can be classified into 
three different types~\cite{BPS1993}:\ 
the first type is 
\begin{subnumcases}{\label{DStype1}}
\mathrm{i} q_{t} + q_{xx} - q_{yy} + \varphi q =0, 
\label{DS1_1}
\\
\varphi_{xx} +\varphi_{yy} = 2 \sigma \left[ \left( \left| q \right|^2 \right)_{xx} - \left( \left| q \right|^2 \right)_{yy} \right], 
\label{DS1_2}
\end{subnumcases}
where \mbox{$\sigma=1$} (focusing case) or \mbox{$\sigma=-1$} (defocusing case); 
the second type is 
\begin{subnumcases}{\label{DStype2}}
\mathrm{i} q_{t} + q_{xx} + q_{yy} + \varphi q =0, 
\label{DS2_1}
\\
\varphi_{xx} -\varphi_{yy} = 2 \left[ \left( \left| q \right|^2 \right)_{xx} + \left( \left| q \right|^2 \right)_{yy} \right];
\label{DS2_2}
\end{subnumcases}
the third type is 
\begin{subnumcases}{\label{DStype3}}
\mathrm{i} q_{t} + q_{xx} - q_{yy} + \varphi q =0, 
\label{DS3_1}
\\
\varphi_{xy} = 2 \left[ \left( \left| q \right|^2 \right)_{xx} - \left( \left| q \right|^2 \right)_{yy} \right]. 
\label{DS3_2}
\end{subnumcases}
Here, the subscripts $t$, $x$ and $y$ denote the partial differentiation with respect to these variables, 
$q$ is a complex-valued function and $\varphi$ is a real-valued function. 
The first type (\ref{DStype1}) 
is outside the scope of this paper, but we stress that only this type 
has two essentially different versions:\ the focusing case (\mbox{$\sigma=1$}) and 
the defocusing case (\mbox{$\sigma=-1$}). 

The second type (\ref{DStype2}) 
(up to 
rotation of the spatial plane) 
and the third type  (\ref{DStype3}) 
can be 
obtained from the system~\cite{Nizh82}:
\begin{subnumcases}{\label{continuousDS}}
\mathrm{i} q_{t} + a \left( q_{xx} + 2F q \right) + b \left( q_{yy} + 2G q \right) =0, 
\label{cDS_1}
\\
F_y = (|q|^2)_x, 
\label{cDS_2}
\\
G_x = (|q|^2)_y, 
\label{cDS_3}
\end{subnumcases}
by 
setting 
\mbox{$a=b=1$} and \mbox{$a=-b=1$}, respectively. 
In (\ref{continuousDS}), 
$a$ and $b$ are real constants,  
and $F$ (defined 
for the case 
\mbox{$a \neq 0$}) and $G$ (defined 
for the case \mbox{$b \neq 0$}) are nonlocal 
real-valued 
potentials, 
so the ``constants'' of integration 
that arise in solving (\ref{cDS_2}) for $F$ and (\ref{cDS_3}) for $G$ 
should also be real-valued. 

Clearly, 
the Davey--Stewartson system 
expressed in the form (\ref{continuousDS})
is a linear combination 
of 
the two elementary flows: 
\begin{equation}
\mathrm{i} q_{t_1} + q_{xx} + 2F q =0, \hspace{5mm} F_y= (|q|^2)_x,
\label{element1}
\end{equation}
and 
\begin{equation}
\mathrm{i} q_{t_2} + q_{yy} + 2G q =0, \hspace{5mm} G_x = (|q|^2)_y.
\label{element2}
\end{equation} 
As described in 
our previous paper~\cite{Tsuchida19}, 
the two elementary 
flows (\ref{element1}) and (\ref{element2}) commute~\cite{Fokas94,Kaji90}; 
that is, 
the relation \mbox{$q_{t_1 t_2} = q_{t_2 t_1}$} 
holds true 
if and only if 
the 
``constants'' of integration appearing 
in $F_{t_2}$ and $G_{t_1}$ 
are chosen appropriately. 
By applying 
a linear change of 
the independent 
variables (see, {\em e.g.}, page 135 of 
\cite{Kono92}): 
\begin{equation}
\widetilde{t} = t_1 + y, \hspace{5mm} \widetilde{x}=x, \hspace{5mm} \widetilde{y}= \alpha y, 
\label{Galilean-like}
\end{equation}
with 
a real constant $\alpha$
to 
(\ref{element1}), 
we obtain 
a \mbox{$(2+1)$}-dimensional generalization~\cite{Mel83} 
of 
the Yajima--Oikawa system~\cite{YO76}:
\begin{equation}
\mathrm{i} q_{t} + q_{xx} + u q =0, \hspace{5mm} u_t + \alpha u_y= 2 (|q|^2)_x. 
\label{2DYO}
\end{equation}
Here, \mbox{$u:=2F$} and 
we omit the tilde 
for brevity. 

The organization of this paper 
is as follows. 
In section 2, we 
provide 
an 
integrable semi-discretization 
(discretization of 
one of the two spatial variables, 
herein $x$) 
of the 
elementary Davey--Stewartson flow 
(\ref{element1}) 
by presenting 
its Lax-pair representation 
and show that it 
admits a straightforward vector generalization. 
By changing the time part of the Lax-pair representation appropriately, 
we 
obtain an integrable semi-discretization 
of the elementary Davey--Stewartson flow (\ref{element2}). 
In section 3, 
we prove 
that 
the two elementary Davey--Stewartson flows 
in the semi-discrete case 
commute under a 
natural choice of 
the ``constants'' of integration. 
Thus, by taking 
a linear combination of the 
semi-discrete elementary Davey--Stewartson flows, 
we arrive at 
an integrable semi-discretization of the 
Davey--Stewartson system (\ref{continuousDS}). 
In addition, 
using 
a linear 
transformation 
of the independent variables 
like 
(\ref{Galilean-like}), 
we can 
convert 
the integrable semi-discretization of 
the elementary Davey--Stewartson flow 
(\ref{element1}) 
to an integrable semi-discretization of 
the \mbox{$(2+1)$}-dimensional Yajima--Oikawa system (\ref{2DYO}), 
which essentially coincides with  
the 
system 
recently 
proposed by G.-F.\ Yu and Z.-W.\ Xu~\cite{Yu15}. 
Note that 
its $y$-independent reduction, i.e., 
the \mbox{$(1+1)$}-dimensional 
discrete 
Yajima--Oikawa system was studied in~\cite{Maruno16, Tsuchida18-1}. 
Concluding remarks are given in 
section~4.

\section{
Integrable semi-discretizations of 
the 
two 
elementary Davey--Stewartson flows}

\subsection{Semi-discrete linear 
problem} 

Inspired by the Lax-pair representation for the \mbox{$(1+1)$}-dimensional 
discrete Yajima--Oikawa system (see Proposition~2.1 in~\cite{Tsuchida18-1}), we consider 
the 
following 
semi-discrete linear problem: 
\begin{subnumcases}{\label{dDS2-L}}
\psi_{n,y} = q_n \phi_n + \chi_n r_n, 
\label{dDS2-L1}
\\[1pt]
\phi_{n+1} -\phi_n = \frac{1}{2} r_{n} \left( \psi_{n+1} + \psi_{n-1} \right),
\label{dDS2-L2}
\\[1pt]
\chi_{n+1} +\chi_n = \frac{1}{2} q_{n}  \left( \psi_{n+1} + \psi_{n-1} \right),
\label{dDS2-L3}
\end{subnumcases}
where $n$ is a discrete spatial variable, 
$y$ is a continuous spatial variable 
and the subscript $y$ denotes the differentiation 
by $y$. 

The linear wavefunction is composed 
of 
three components $\psi_n$, $\phi_n$ and $\chi_n$; 
this is in contrast with the 
two-component 
wavefunction satisfying the 
spatial linear problem 
for 
the 
continuous Davey--Stewartson system (\ref{continuousDS})~\cite{Hab75,Nizh82}: 
\begin{subnumcases}{\label{clinear_s}}
\psi_{y} = q \phi, 
\label{clinear_s1}
\\
\phi_{x} = -q^\ast \psi,
\label{clinear_s2}
\end{subnumcases}
where the asterisk denotes the 
complex 
conjugation. 

The dependent variables $q_n$ and $r_n$ in (\ref{dDS2-L}) 
are scalars, 
but 
it is 
possible to consider a more general case of 
vector-valued variables $\vt{q}_n$ and $\vt{r}_n$, 
which 
will be touched upon at the end of subsection~\ref{subsec2.2}. 
Note that 
the first equation (\ref{dDS2-L1}) can be 
rewritten as 
\begin{equation}
\nonumber
\psi_{n,y}= q_n \phi_{n+1} - \chi_{n+1} r_n, 
\end{equation}
using the second and third equations (\ref{dDS2-L2}) and (\ref{dDS2-L3}).

\subsection{Semi-discretization 
of the elementary Davey--Stewartson flow (\ref{element1})}
\label{subsec2.2}

In view of the time part of 
the Lax-pair representation for the \mbox{$(1+1)$}-dimensional 
discrete Yajima--Oikawa system (see Proposition~2.1 in~\cite{Tsuchida18-1}), we consider 
the following time evolution of the linear wavefunction: 
\begin{subnumcases}{\label{dDS2-M1}}
\mathrm{i} \psi_{n,t_1} = v_{n} \left( \psi_{n+1} + \psi_{n-1} \right) - c \psi_n, 
\label{dDS2-M11}
\\[1pt]
\mathrm{i} \phi_{n,t_1} = \frac{1}{2} v_n r_{n-1} \left( \psi_{n+1} + \psi_{n-1} \right) 
 - \frac{1}{2} v_{n-1} r_{n} \left( \psi_{n} + \psi_{n-2} \right), 
\\[1pt]
\mathrm{i} \chi_{n,t_1} = \frac{1}{2} v_n q_{n-1} \left( \psi_{n+1} + \psi_{n-1} \right) 
 + \frac{1}{2} v_{n-1} q_{n} \left( \psi_{n} + \psi_{n-2} \right) - 2c \chi_n, \;\;
\label{dDS2-M13}
\end{subnumcases}
%
where $c$ is an arbitrary constant and $v_n$ is 
a scalar auxiliary 
function. 


\begin{proposition}
\label{prop2.1}
The compatibility 
conditions of the 
overdetermined 
linear 
systems 
\mbox{$(\ref{dDS2-L})$} and \mbox{$(\ref{dDS2-M1})$} 
for $\psi_n$, $\phi_n$ and $\chi_n$ 
are equivalent to the following semi-discrete 
system in \mbox{$2+1$} dimensions:  
\begin{subnumcases}{\label{dDS1-f1}}
\mathrm{i} q_{n,t_1} = v_n \left( q_{n+1}+q_{n-1} \right)  - c \hspace{1pt} q_{n}, 
\label{} \\[1pt]
\mathrm{i} r_{n,t_1} = -v_n \left( r_{n+1}+r_{n-1} \right) + c \hspace{1pt} r_{n}, 
\label{} \\[1pt]
v_{n,y} = \frac{1}{2} v_n 
\left( q_{n} r_{n-1} + q_{n-1} r_{n} -  q_{n+1} r_{n} - q_{n} r_{n+1} \right). 
\label{} 
\end{subnumcases}
\end{proposition}
We can prove this 
proposition 
by a direct calculation. 
Specifically, 
using (\ref{dDS2-L}) and (\ref{dDS2-M1}),  
the compatibility conditions can be rewritten as 
\begin{align}
0 & = \mathrm{i} \psi_{n,y t_1} - \mathrm{i} \psi_{n,t_1 y} 
\nonumber \\
& = \left( \mathrm{i} q_{n,t_1} - v_n q_{n+1} - v_n q_{n-1} +c q_n \right) \phi_n 
	+ \chi_n \left( \mathrm{i} r_{n,t_1} + v_n r_{n+1} + v_n r_{n-1} - c r_n \right) 
\nonumber \\
& \hphantom{=} \; \, \mbox{}
	+ \left[ -v_{n,y} + \frac{1}{2} v_n  \left( q_{n} r_{n-1} + q_{n-1} r_{n} -  q_{n+1} r_{n} - q_{n} r_{n+1} \right) \right] 
	\left( \psi_{n+1} + \psi_{n-1} \right), 
\nonumber 
\end{align}
\begin{align}
0 &= \mathrm{i} \left[ \frac{1}{2} r_{n} \left( \psi_{n+1} + \psi_{n-1} \right)- \phi_{n+1}  + \phi_{n} \right]_{t_1} 
\nonumber \\
& = \frac{1}{2} \left( \mathrm{i} r_{n,t_1} + v_n r_{n+1} + v_n r_{n-1} -c r_n \right) \left( \psi_{n+1} + \psi_{n-1} \right),
\nonumber 
\end{align}
and 
\begin{align}
0 &= \mathrm{i} \left[ \frac{1}{2} q_{n} \left( \psi_{n+1} + \psi_{n-1} \right) - \chi_{n+1} - \chi_{n} \right]_{t_1} 
\nonumber \\
& = \frac{1}{2} \left( \mathrm{i} q_{n,t_1} - v_n q_{n+1} - v_n q_{n-1} + c q_n \right) \left( \psi_{n+1} + \psi_{n-1} \right),
\nonumber 
\end{align}
which 
imply 
(\ref{dDS1-f1}) and vice versa. 

If \mbox{$c \in \mathbb{R}$}, 
we can impose 
the complex conjugation reduction: 
\begin{equation}
r_n =  - \varDelta q_n^\ast, \hspace{5mm} 
v_n^\ast=v_n, 
\nonumber
\end{equation}
on the system (\ref{dDS1-f1}), 
where 
$\varDelta$ is 
an arbitrary real constant. 
In particular, if we set 
\begin{align}
v_n = \frac{1}{\varDelta^2} + F_n, \hspace{5mm} 
c = \frac{2}{\varDelta^2}, 
\nonumber 
\end{align}
in (\ref{dDS1-f1}) 
and impose the complex conjugation reduction, 
we obtain 
\begin{subnumcases}{\label{rdDS1-f1}}
\mathrm{i} q_{n,t_1} =  \frac{1}{\varDelta^2} \left( q_{n+1}+q_{n-1} -2 q_{n} \right) + F_n \left( q_{n+1}+q_{n-1} \right), 
\label{} \\[1pt]
F_{n,y} = \frac{1}{2} \left( \frac{1}{\varDelta} + \varDelta F_n \right) 
\left( q_{n+1} q_{n}^\ast + q_{n} q_{n+1}^\ast - q_{n} q_{n-1}^\ast - q_{n-1} q_{n}^\ast \right),
\label{} 
\end{subnumcases}
where \mbox{$q_n \in \mathbb{C}$} and \mbox{$F_n \in \mathbb{R}$}. 


By further 
setting 
\begin{align}
 q_n(y, t_1) = q (n \varDelta, y, t_1), \hspace{5mm} 
F_n (y,t_1) = F (n \varDelta, y, t_1),
\nonumber 
\end{align}
the semi-discrete system (\ref{rdDS1-f1}) 
reduces in the continuous limit \mbox{$\varDelta \to 0$} 
to 
\begin{equation} 
\label{reduced_first_flow4}
\left\{ 
\begin{split}
& \mathrm{i} q_{t_1} 
= q_{xx} + 2 F q, 
\\[2pt]
& F_{y} = \left( \left| q \right|^2 \right)_x, 
\end{split} 
\right. 
\end{equation}
where \mbox{$x:= n \Delta$}. 
The system 
(\ref{reduced_first_flow4})
can be 
identified with the elementary Davey--Stewartson flow (\ref{element1}), 
up to a sign inversion of $t_1$. 
Thus, 
(\ref{rdDS1-f1}) 
can be interpreted 
as 
an integrable semi-discretization 
of 
the elementary Davey--Stewartson flow 
(\ref{element1}). 

As in the continuous case~\cite{Kono92,Calogero91,
Fordy87,March}, 
the semi-discrete 
system (\ref{dDS1-f1}) 
admits a straightforward vector generalization:
\begin{subnumcases}{\label{dvDS1-f1}}
\mathrm{i} \vt{q}_{n,t_1} = v_n \left( \vt{q}_{n+1}+\vt{q}_{n-1} \right)  - c \hspace{1pt} \vt{q}_{n}, 
\label{} \\[1pt]
\mathrm{i} \vt{r}_{n,t_1} = -v_n \left( \vt{r}_{n+1}+\vt{r}_{n-1} \right) + c \hspace{1pt} \vt{r}_{n}, 
\label{} \\[1pt]
v_{n,y} = \frac{1}{2} v_n 
\left( \sca{\vt{q}_{n}}{\vt{r}_{n-1}} + \sca{\vt{q}_{n-1}}{\vt{r}_{n}} -  \sca{\vt{q}_{n+1}}{\vt{r}_{n}} - \sca{\vt{q}_{n}}{\vt{r}_{n+1}} \right). \;\;\;
\label{} 
\end{subnumcases}
Here, \mbox{$\sca{\,\cdot\,}{\,\cdot\,}$} 
stands for 
the standard 
scalar product. 
The Lax-pair representation for (\ref{dvDS1-f1})
is given by the overdetermined linear systems: 
\begin{equation}
\label{vectorDS2-L}
\left\{ 
\begin{split}
& \psi_{n,y} = \sca{\vt{q}_n}{\vt{\phi}_n} + \sca{\vt{\chi}_n}{\vt{r}_{n}}, 
\\[1pt]
& \vt{\phi}_{n+1} -\vt{\phi}_n = \frac{1}{2} \vt{r}_{n} \left( \psi_{n+1} + \psi_{n-1} \right),
\\[1pt]
& \vt{\chi}_{n+1} +\vt{\chi}_n = \frac{1}{2}  \vt{q}_{n}  \left( \psi_{n+1} + \psi_{n-1} \right),
\end{split}
\right.
\end{equation}
and 
\begin{equation}
\nonumber
\left\{ 
\begin{split}
& \mathrm{i} \psi_{n,t_1} = v_{n} \left( \psi_{n+1} + \psi_{n-1} \right) - c \psi_n, 
\\[1pt]
& \mathrm{i} \vt{\phi}_{n,t_1} = \frac{1}{2} v_n \vt{r}_{n-1} \left( \psi_{n+1} + \psi_{n-1} \right) 
 - \frac{1}{2} v_{n-1} \vt{r}_{n} \left( \psi_{n} + \psi_{n-2} \right), 
\\[1pt]
& \mathrm{i} \vt{\chi}_{n,t_1} = \frac{1}{2} v_n \vt{q}_{n-1} \left( \psi_{n+1} + \psi_{n-1} \right) 
 + \frac{1}{2} v_{n-1} \vt{q}_{n} \left( \psi_{n} + \psi_{n-2} \right) - 2c \vt{\chi}_n. \;\;
\end{split}
\right.
\end{equation}
Note that the \mbox{$(1+1)$}-dimensional (\mbox{$\partial_{t_1}=\partial_y$})
reduction of 
the system (\ref{dvDS1-f1})
was studied in~\cite{Maruno16, Tsuchida18-1}.

\subsection{Semi-discretization
of the elementary Davey--Stewartson flow (\ref{element2})}

To obtain an integrable 
semi-discretization of the elementary Davey--Stewartson flow (\ref{element2}), 
we consider 
the following 
time evolution of the linear wavefunction involving 
differentiation 
with respect to the continuous spatial variable $y$: 
\begin{subnumcases}{\label{sd_time2}}
\mathrm{i} \psi_{n,t_{2}} = -q_n \phi_{n,y} + \left( q_{n,y} - q_n q_{n-1} r_n \right) \phi_n
\nonumber \\ \hspace{15mm} \mbox{} 
	+ r_n \chi_{n,y} - \left( r_{n,y} - q_n r_n r_{n-1} \right) \chi_n,
\label{sd_time2_1}
\\[4pt]
\mathrm{i} \phi_{n,t_{2}} =  -\phi_{n,yy}  - \left[ G_n + \left( q_{n-1} r_{n} \right)_y \right] \phi_{n} + r_n r_{n-1} \chi_{n,y} + J_n \chi_n, 
\label{sd_time2_2}
\\[1pt]
\mathrm{i} \chi_{n,t_{2}} =  -q_n q_{n-1} \phi_{n,y} + H_n \phi_n + \chi_{n,yy} + \left[ G_n + \left( q_n r_{n-1} \right)_y \right] \chi_{n}, 
\label{sd_time2_3}
\end{subnumcases}
where $G_n$, $H_n$ and $J_n$ are scalar auxiliary functions. 

\begin{proposition}
\label{}
The compatibility 
conditions of the 
overdetermined 
linear 
systems 
\mbox{$(\ref{dDS2-L})$} and \mbox{$(\ref{sd_time2})$} 
for $\psi_n$, $\phi_n$ and $\chi_n$
are equivalent to the following semi-discrete 
system in \mbox{$2+1$} dimensions:  
\begin{equation} 
\label{second_flow}
\left\{ 
\begin{split}
& \mathrm{i} q_{n,t_{2}} = q_{n,yy} + G_n q_n - H_n r_n - q_{n,y} q_{n-1} r_n, 
\\[2pt]
& \mathrm{i} r_{n,t_{2}} = - r_{n,yy} - J_n q_n - G_n r_n + q_n r_{n-1} r_{n,y}, 
\\[2pt]
& G_{n+1} - G_{n} = - \left( q_{n+1} r_n + q_n r_{n+1} \right)_y + \frac{1}{2} q_n r_n \left( q_{n+1} r_{n+1} -q_{n-1} r_{n-1} \right), 
\\[2pt]
& H_{n+1} + H_{n} = -q_{n,y} \left( q_{n+1} + q_{n-1} \right) - \frac{1}{2} q_n^2 \left( q_{n+1} r_{n+1} -q_{n-1} r_{n-1} \right), 
\\[2pt]
& J_{n+1} + J_{n} = r_{n,y} \left( r_{n+1} + r_{n-1} \right) + \frac{1}{2} r_n^2 \left( q_{n+1} r_{n+1} -q_{n-1} r_{n-1} \right). 
\end{split} 
\right. 
\end{equation}
\end{proposition}

This proposition can be proved by a straightforward calculation. 
Indeed, 
using 
(\ref{dDS2-L}) and (\ref{sd_time2}),  we 
can rewrite the compatibility 
conditions as 
\begin{align}
0 & = \mathrm{i} \left( \psi_{n,y t_2} - \psi_{n,t_2 y} \right)
\nonumber \\
& = \left( \mathrm{i} q_{n,t_{2}} - q_{n,yy} -G_n q_n +H_n r_n + q_{n,y} q_{n-1} r_n \right) \phi_n 
\nonumber \\
& \hphantom{=} \; \, \mbox{}
	+ \left( \mathrm{i} r_{n,t_{2}} + r_{n,yy} + J_n q_n +G_n r_n - q_{n} r_{n-1} r_{n,y} \right) \chi_n ,  
\nonumber 
\end{align}
\begin{align}
0 &=  \mathrm{i} \left[ \frac{1}{2} r_{n} \left( \psi_{n+1} + \psi_{n-1} \right) + \phi_{n} - \phi_{n+1} \right]_{t_2} 
\nonumber \\
& =  \frac{1}{2} \left[ \mathrm{i} r_{n,t_{2}} + r_{n,yy} - J_{n+1} q_n +G_{n+1} r_n +\left( q_{n+1} r_{n} \right)_y r_{n} 
	+ \left( q_{n} r_{n+1} r_{n} \right)_y \right]
	\left( \psi_{n+1} + \psi_{n-1} \right)
\nonumber \\
& \hphantom{=} \; \, \mbox{}
	+ \left[ G_{n+1} - G_{n} + \left( q_{n+1} r_n + q_n r_{n+1} \right)_y - \frac{1}{2} q_n r_n \left( q_{n+1} r_{n+1} -q_{n-1} r_{n-1} \right)
 	\right] \phi_n
\nonumber \\
& \hphantom{=} \; \, \mbox{}
	+ \left[ J_{n+1} + J_{n} - r_{n,y} \left( r_{n+1} + r_{n-1} \right) - \frac{1}{2} r_n^2 \left( q_{n+1} r_{n+1} -q_{n-1} r_{n-1} \right)
	\right] \chi_n,
\nonumber 
\end{align}
and 
\begin{align}
0 &= \mathrm{i} \left[ \frac{1}{2} q_{n} \left( \psi_{n+1} + \psi_{n-1} \right) - \chi_{n} - \chi_{n+1} \right]_{t_2} 
\nonumber \\
& =  \frac{1}{2} \left[ \mathrm{i} q_{n,t_{2}} - q_{n,yy} - G_{n+1} q_n - H_{n+1} r_n -\left( q_{n} r_{n+1} \right)_y q_{n} 
	- \left( q_{n+1} q_{n} r_{n} \right)_y \right]
	\left( \psi_{n+1} + \psi_{n-1} \right)
\nonumber \\
& \hphantom{=} \; \, \mbox{}
	- \left[  H_{n+1} + H_{n} + q_{n,y} \left( q_{n+1} + q_{n-1} \right) + \frac{1}{2} q_n^2 \left( q_{n+1} r_{n+1} -q_{n-1} r_{n-1} \right) 
	\right] \phi_n
\nonumber \\
& \hphantom{=} \; \, \mbox{}
	+ \left[  G_{n+1} - G_{n} + \left( q_{n+1} r_n + q_n r_{n+1} \right)_y - \frac{1}{2} q_n r_n \left( q_{n+1} r_{n+1} -q_{n-1} r_{n-1} \right)
	\right] \chi_n,
\nonumber 
\end{align}
which 
are,  as a whole, 
equivalent to 
(\ref{second_flow}). 

We can impose the complex conjugation reduction 
\mbox{$r_n= - \varDelta q_n^\ast$}, \mbox{$G_n^\ast = G_n$} 
and \mbox{$J_n = - \varDelta^2 H_n^\ast$}
on (\ref{second_flow}) 
to 
obtain 
\begin{equation} 
\label{second_flow_cc}
\left\{ 
\begin{split}
& \mathrm{i} q_{n,t_{2}} = q_{n,yy} + G_n q_n + \varDelta H_n q_n^\ast + \varDelta q_{n,y} q_{n-1} q_n^\ast, 
\\[2pt]
& G_{n+1} - G_{n} = \varDelta \left( q_{n+1} q_n^\ast + q_n q_{n+1}^\ast \right)_y 
	+ \frac{1}{2} \varDelta^2 \left| q_n \right|^2 \left( \left| q_{n+1} \right|^2 - \left| q_{n-1} \right|^2 \right), 
\\[2pt]
& H_{n+1} + H_{n} = -q_{n,y} \left( q_{n+1} + q_{n-1} \right) + \frac{1}{2}  \varDelta q_n^2 \left( \left| q_{n+1} \right|^2
	- \left| q_{n-1} \right|^2 \right),
\end{split} 
\right. 
\end{equation}
where 
$\varDelta$ is 
an arbitrary real constant. 
If we interpret $\varDelta$ as a lattice parameter, 
set \mbox{$x:= n \Delta$} 
and consider 
the continuous limit \mbox{$\varDelta \to 0$}, 
(\ref{second_flow_cc}) 
reduces to the elementary Davey--Stewartson flow (\ref{element2}), 
up to time reversal 
and a minor 
change of notation. 

In subsection~\ref{subsec2.2}, 
we showed that 
the semi-discrete 
system (\ref{dDS1-f1}) 
admits 
the vector generalization
(\ref{dvDS1-f1}). 
Analogously, 
we can construct a vector generalization 
of the semi-discrete 
system (\ref{second_flow}), 
which is associated with the linear 
problem 
(\ref{vectorDS2-L}). 
However, the equations of motion for 
this vector generalization 
are highly nonlocal and 
complicated, 
so we do not present them here.

\section{Integrable semi-discretizations of 
the 
Davey--Stewartson 
system 
and 
the 
\mbox{$(2+1)$}-dimensional Yajima--Oikawa system} 

Let us first demonstrate that 
in the generic case 
the semi-discrete 
flow 
(\ref{dDS1-f1}) and the semi-discrete 
flow
(\ref{second_flow}) commute.  
Using (\ref{dDS1-f1}) and (\ref{second_flow}), 
we obtain 
\begin{align}
2 \mathrm{i} \left( \log v_{n} \right)_{y t_2} 
	&= \mathrm{i} \left( q_{n} r_{n-1} + q_{n-1} r_{n} - q_{n+1} r_{n} - q_{n} r_{n+1} \right)_{t_2}
\nonumber \\
&= \left[ q_{n,y} r_{n-1} - q_{n} r_{n-1,y} + q_{n-1,y} r_{n} - q_{n-1} r_{n,y} -\frac{1}{2} \left( q_n r_{n-1} \right)^2 
	+ \frac{1}{2} \left( q_{n-1} r_{n} \right)^2 \right]_y 
\nonumber \\
& \hphantom{=} \; \, \mbox{ } - \left[ n \to n+1 \right]_y, 
\nonumber
\end{align}
which implies
\begin{align}
 \mathrm{i} \left( \log v_{n} \right)_{t_2} 
	&= \mathrm{i} \frac{v_{n,t_2}}{v_n} 
\nonumber \\
&= \left[ \frac{1}{2} \left( q_{n,y} r_{n-1} - q_{n} r_{n-1,y} + q_{n-1,y} r_{n} - q_{n-1} r_{n,y} \right) -\frac{1}{4} \left( q_n r_{n-1} \right)^2 
	+ \frac{1}{4} \left( q_{n-1} r_{n} \right)^2 \right]
\nonumber \\
& \hphantom{=} \; \, \mbox{ } - \left[ n \to n+1 \right] + f_n (t_1, t_2), 
\label{v_t_2}
\end{align}
where $f_n (t_1, t_2)$ is a $y$-independent 
function. 
Moreover, using 
(\ref{dDS1-f1}) and (\ref{second_flow}), 
we also 
obtain 
\begin{align}
 \mathrm{i} G_{n, t_1} 
	&= \frac{1}{2} v_n \left[ -\left( q_{n} r_{n-1} - q_{n+1} r_{n} - q_{n} r_{n+1} \right) q_{n+1} r_{n-1} 
	+ \left( q_{n-1} r_{n} - q_{n+1} r_{n} - q_{n} r_{n+1} \right) q_{n-1} r_{n+1} \right. 
\nonumber \\
& \hphantom{=} \, \left. \mbox{} + q_{n-1} r_{n-1} \left( q_{n-1} r_{n} - q_{n} r_{n-1} \right) \right]
	-v_n \left( q_{n+1} r_{n-1} - q_{n-1} r_{n+1} \right)_y 
\nonumber \\
& \hphantom{=} \; \, \mbox{} + \frac{1}{2} v_{n-1} q_{n} r_{n} \left( q_{n} r_{n-1} + q_{n-2} r_{n-1} - q_{n-1} r_{n} - q_{n-1} r_{n-2} \right)
	+ g (y, t_1, t_2), 
\label{G_t_1}
\end{align}
\begin{align}
 \mathrm{i} H_{n, t_1} 
	&= -2c H_n - v_{n-1} q_{n,y} \left( q_n + q_{n-2} \right) - v_n q_{n-1,y} \left( q_{n+1} + q_{n-1} \right)
\nonumber \\
& \hphantom{=} \; \, \mbox{} -\frac{1}{2} v_{n} q_{n-1}^{\,2} \left( q_{n+1} r_{n} + q_{n-1} r_{n} - q_{n} r_{n+1} -q_{n} r_{n-1} \right)  
\nonumber \\
& \hphantom{=} \; \, \mbox{} + \frac{1}{2} v_{n-1} q_{n}^2 \left( q_{n} r_{n-1} + q_{n-2} r_{n-1} - q_{n-1} r_{n} - q_{n-1} r_{n-2} \right)
	+ \left( -1 \right) ^n h (y, t_1, t_2), 
\label{H_t_1}
\end{align}
and 
\begin{align}
 \mathrm{i} J_{n, t_1} 
	&= 2c J_n - v_{n} r_{n-1,y} \left( r_{n+1} + r_{n-1} \right) - v_{n-1} r_{n,y} \left( r_{n} + r_{n-2} \right)
\nonumber \\
& \hphantom{=} \; \, \mbox{} +\frac{1}{2} v_{n} r_{n-1}^{\,2} \left( q_{n+1} r_{n} + q_{n-1} r_{n} - q_{n} r_{n+1} -q_{n} r_{n-1} \right)  
\nonumber \\
& \hphantom{=} \; \, \mbox{} - \frac{1}{2} v_{n-1} r_{n}^2 \left( q_{n} r_{n-1} + q_{n-2} r_{n-1} - q_{n-1} r_{n} - q_{n-1} r_{n-2} \right)
	+ \left( -1 \right) ^n j (y, t_1, t_2), 
\label{J_t_1}
\end{align}
where $g (y, t_1, t_2)$, 
$h(y, t_1, t_2)$ and $j(y, t_1, t_2)$ are 
$n$-independent 
functions. 

By a straightforward 
calculation, we can prove 
the following proposition. 
\begin{proposition}
\label{}
Equations 
\mbox{$(\ref{dDS1-f1})$}, \mbox{$(\ref{second_flow})$} 
and 
\mbox{$(\ref{v_t_2})$}--\mbox{$(\ref{J_t_1})$}
imply 
the commutativity of $\partial_{t_1}$ and $\partial_{t_2}$, 
i.e., 
\begin{equation}
\nonumber 
q_{n, t_1 t_2} = q_{n, t_2 t_1} \;\, \mathrm{and} \;\, r_{n, t_1 t_2} = r_{n, t_2 t_1}, 
\end{equation}
if and only if  the ``constants'' of integration $f_n (t_1, t_2)$, 
$g (y, t_1, t_2)$, $h(y, t_1, t_2)$ and $j(y, t_1, t_2)$ all vanish identically. 
\end{proposition}

Note that 
it is possible to decompose the semi-discrete 
flow (\ref{dDS1-f1}) into more fundamental flows 
by extracting the trivial zeroth flow from (\ref{dDS1-f1}): 
\begin{subnumcases}{\label{zeroth_flow}}
q_{n,t_0} = -q_{n}, 
\label{} \\
r_{n,t_0} = r_{n}. 
\label{} 
\end{subnumcases}
In the generic case, 
the zeroth flow (\ref{zeroth_flow}) 
commutes with 
the semi-discrete 
flow 
(\ref{dDS1-f1}) (for any value of $c$, say, \mbox{$c=0$}) and the semi-discrete 
flow
(\ref{second_flow}); 
that is, 
\begin{equation}
\nonumber 
q_{n, t_0 t_1} = q_{n, t_1 t_0} \;\, \mathrm{and} \;\, r_{n, t_0 t_1} = r_{n, t_1 t_0}, 
\end{equation}
and 
\begin{equation}
\nonumber 
q_{n, t_0 t_2} = q_{n, t_2 t_0} \;\, \mathrm{and} \;\, r_{n, t_0 t_2} = r_{n, t_2 t_0}, 
\end{equation}
if 
the corresponding ``constants'' of integration 
vanish. 

In view of the commutativity of the 
semi-discrete 
flow 
(\ref{dDS1-f1}) and the semi-discrete 
flow
(\ref{second_flow}), 
we can naturally 
consider a linear combination of 
the 
two flows: 
\begin{equation}
\nonumber 
\partial_t := a \partial_{t_1} + b \partial_{t_2}, 
\end{equation}
with the change of notation \mbox{$a c \to \beta$}. 
Thus, the time evolution of the linear wavefunction 
can be written as 
\begin{subnumcases}{\label{sd_time_g}}
\mathrm{i} \psi_{n,t} =a \hspace{1pt} v_{n} \left( \psi_{n+1} + \psi_{n-1} \right) - \beta  \psi_n
	+ b \left[ -q_n \phi_{n,y} + \left( q_{n,y} - q_n q_{n-1} r_n \right) \phi_n \right. 
\nonumber \\ \hspace{10mm} \left. \mbox{} 
	+ r_n \chi_{n,y} - \left( r_{n,y} - q_n r_n r_{n-1} \right) \chi_n \right],
\label{sd_time_g1}
\\[4pt]
\mathrm{i} \phi_{n,t} =  a \left[ \frac{1}{2} v_n r_{n-1} \left( \psi_{n+1} + \psi_{n-1} \right) 
 - \frac{1}{2} v_{n-1} r_{n} \left( \psi_{n} + \psi_{n-2} \right) \right] 
\nonumber \\ \hspace{10mm} \mbox{} 
	+ b \left\{ -\phi_{n,yy}  - \left[ G_n + \left( q_{n-1} r_{n} \right)_y \right] \phi_{n} + r_n r_{n-1} \chi_{n,y} 
	+ J_n \chi_n \right\}, 
\label{sd_time_g2}
\\[1pt]
\mathrm{i} \chi_{n,t} = a \left[ \frac{1}{2} v_n q_{n-1} \left( \psi_{n+1} + \psi_{n-1} \right) 
 + \frac{1}{2} v_{n-1} q_{n} \left( \psi_{n} + \psi_{n-2} \right) \right] - 2\beta \chi_n
\nonumber \\ \hspace{10mm}  \mbox{} + b \left\{
	-q_n q_{n-1} \phi_{n,y} + H_n \phi_n + \chi_{n,yy} + \left[ G_n + \left( q_n r_{n-1} \right)_y \right] \chi_{n} \right\}, 
\label{sd_time_g3}
\end{subnumcases}
where
$a$, $\beta$ and $b$ are 
constants (or, more generally, arbitrary functions of the time variable $t$~\cite{Calo76}) 
and $v_n$, 
$G_n$, $H_n$ and $J_n$ are 
auxiliary functions. 

By a 
straightforward calculation, we can 
prove 
that the following proposition holds true. 
\begin{proposition}
\label{prop3.2}
The compatibility 
conditions of the 
overdetermined 
linear 
systems 
\mbox{$(\ref{dDS2-L})$} and \mbox{$(\ref{sd_time_g})$} 
for $\psi_n$, $\phi_n$ and $\chi_n$ 
are equivalent to the following semi-discrete 
system in \mbox{$2+1$} dimensions:   
\begin{equation} 
\label{first+second_flow}
\left\{ 
\begin{split}
& \mathrm{i} q_{n,t} = a \hspace{1pt} v_n \left( q_{n+1}+q_{n-1} \right)  - \beta \hspace{1pt} q_{n} 
	+ b \left( q_{n,yy} + G_n q_n - H_n r_n - q_{n,y} q_{n-1} r_n \right), 
\\[2pt]
& \mathrm{i} r_{n,t} = -a \hspace{1pt} v_n \left( r_{n+1}+r_{n-1} \right) + \beta \hspace{1pt} r_{n}
	+ b \left( - r_{n,yy} - J_n q_n - G_n r_n + q_n r_{n-1} r_{n,y} \right), 
\\[2pt]
& v_{n,y} = \frac{1}{2} v_n 
\left( q_{n} r_{n-1} + q_{n-1} r_{n} -  q_{n+1} r_{n} - q_{n} r_{n+1} \right) \hspace{3mm} \mathrm{if} \; 
	a \neq 0,
\\[2pt]
& G_{n+1} - G_{n} = - \left( q_{n+1} r_n + q_n r_{n+1} \right)_y + \frac{1}{2} q_n r_n \left( q_{n+1} r_{n+1} -q_{n-1} r_{n-1} \right) 
\hspace{3mm} \mathrm{if} \; b \neq 0,
\\[2pt]
& H_{n+1} + H_{n} = -q_{n,y} \left( q_{n+1} + q_{n-1} \right) - \frac{1}{2} q_n^2 \left( q_{n+1} r_{n+1} -q_{n-1} r_{n-1} \right)
\hspace{3mm} \mathrm{if} \; b \neq 0, 
\\[2pt]
& J_{n+1} + J_{n} = r_{n,y} \left( r_{n+1} + r_{n-1} \right) + \frac{1}{2} r_n^2 \left( q_{n+1} r_{n+1} -q_{n-1} r_{n-1} \right)
\hspace{3mm} \mathrm{if} \; b \neq 0. 
\end{split} 
\right. 
\end{equation}
\end{proposition}

If 
\mbox{$a, \beta, b \in \mathbb{R}$}, 
we can impose 
the complex conjugation reduction: 
\begin{equation}
r_n =  - \varDelta q_n^\ast, \hspace{5mm} v_n^\ast=v_n, \hspace{5mm} 
G_n^\ast = G_n, \hspace{5mm} J_n = - \varDelta^2 H_n^\ast, 
\nonumber
\end{equation}
on the system (\ref{first+second_flow}), 
where 
$\varDelta$ is 
an arbitrary real constant. 
By 
setting 
\begin{align}
v_n = \frac{1}{\varDelta^2} + F_n, \hspace{5mm} 
\beta = \frac{2}{\varDelta^2} a, 
\nonumber 
\end{align}
the complex conjugation reduction simplifies (\ref{first+second_flow}) to 
\begin{equation} 
\label{semi-discrete_DS}
\left\{ 
\begin{split}
& \mathrm{i} q_{n,t} = a \left[ \frac{1}{\varDelta^2} \left( q_{n+1}+q_{n-1} -2 q_{n} \right) 
	+ F_n \left( q_{n+1}+q_{n-1} \right) \right]
\\ 
& \hphantom{\mathrm{i} q_{n,t} =} \mbox{} 
+ b \left( q_{n,yy} + G_n q_n + \varDelta H_n q_n^\ast + \varDelta q_{n,y} q_{n-1} q_n^\ast \right), 
\\[2pt]
& F_{n,y} = \frac{1}{2} \left( \frac{1}{\varDelta} + \varDelta F_n \right)
\left( q_{n+1} q_{n}^\ast + q_{n} q_{n+1}^\ast - q_{n} q_{n-1}^\ast - q_{n-1} q_{n}^\ast \right) \hspace{3mm} \mathrm{if} \; a \neq 0, 
\\[2pt]
& G_{n+1} - G_{n} = \varDelta \left( q_{n+1} q_n^\ast + q_n q_{n+1}^\ast \right)_y 
	+ \frac{1}{2} \varDelta^2 \left| q_n \right|^2 \left( \left| q_{n+1} \right|^2 - \left| q_{n-1} \right|^2 \right) 
\hspace{3mm} \mathrm{if} \; b \neq 0, 
\\[2pt]
& H_{n+1} + H_{n} = -q_{n,y} \left( q_{n+1} + q_{n-1} \right) + \frac{1}{2}  \varDelta q_n^2 \left( \left| q_{n+1} \right|^2
	- \left| q_{n-1} \right|^2 \right) \hspace{3mm} \mathrm{if} \; b \neq 0, 
\end{split} 
\right. 
\end{equation}
where 
$a$ and $b$ are 
real 
constants, 
\mbox{$q_n, H_n \in \mathbb{C}$} and \mbox{$F_n, G_n \in \mathbb{R}$}. 
The semi-discrete system (\ref{semi-discrete_DS}) with a sign inversion of $t$ and 
a minor 
change of notation 
reduces in the continuous limit \mbox{$\varDelta \to 0$} 
to the Davey--Stewartson system (\ref{continuousDS}), 
where \mbox{$x:= n \Delta$}. 
Thus, (\ref{semi-discrete_DS}) 
can be regarded as an integrable semi-discretization 
of the Davey--Stewartson system (\ref{continuousDS}). 

If we 
consider 
a linear change 
of the independent variables: 
\begin{equation}
\widetilde{t} = t_1 + y, \hspace{5mm} \widetilde{y}= \alpha y, 
\label{Galilean-like2}
\end{equation}
where $\alpha$ is an arbitrary real constant, 
the semi-discrete elementary Davey--Stewartson flow (\ref{rdDS1-f1})  
is transformed to 
\begin{equation} 
\label{sd_YO1}
\left\{ 
\begin{split}
& \mathrm{i} q_{n,t} =  \frac{1}{\varDelta^2} \left( q_{n+1}+q_{n-1} -2 q_{n} \right) 
	+ F_n \left( q_{n+1}+q_{n-1} \right), 
\\[1pt]
& F_{n,t} + \alpha F_{n,y} = \frac{1}{2} \left( \frac{1}{\varDelta} + \varDelta F_n \right) 
\left( q_{n+1} q_{n}^\ast + q_{n} q_{n+1}^\ast - q_{n} q_{n-1}^\ast - q_{n-1} q_{n}^\ast \right). 
\end{split} 
\right. 
\end{equation}
Here, we omit 
the tilde of the continuous independent variables 
for brevity.  
The system 
(\ref{sd_YO1}) 
can be interpreted 
as 
an integrable semi-discretization of 
the \mbox{$(2+1)$}-dimensional Yajima--Oikawa system (\ref{2DYO}), 
up to a 
rescaling of variables; 
note that 
(\ref{sd_YO1}) essentially coincides with  
the 
system 
recently 
proposed by G.-F.\ Yu and Z.-W.\ Xu~\cite{Yu15}. 
If we discard the dependence on $y$, 
(\ref{sd_YO1}) 
reduces to the 
\mbox{$(1+1)$}-dimensional 
discrete 
Yajima--Oikawa system 
studied in~\cite{Maruno16, Tsuchida18-1}.

\section{Concluding remarks}

As a continuation of 
our previous paper~\cite{Tsuchida19}, 
we studied 
the problem of 
how to 
discretize the spatial variable $x$ 
in 
the Davey--Stewartson system (\ref{continuousDS}) 
and the \mbox{$(2+1)$}-dimensional Yajima--Oikawa system (\ref{2DYO}). 
To 
guarantee the integrability of the semi-discretization, 
we 
start with the linear problem (\ref{dDS2-L}), 
wherein we can impose the 
complex conjugation reduction 
\mbox{$r_n =  - \varDelta q_n^\ast$} 
with a real constant $\varDelta$. 
By associating (\ref{dDS2-L}) 
with an appropriate 
time-evolutionary system 
of the linear wavefunction 
and computing the compatibility conditions, 
we obtain
(\ref{rdDS1-f1}) 
(resp.~(\ref{second_flow_cc})) 
as 
an integrable semi-discretization 
of the elementary Davey--Stewartson flow (\ref{element1}) (resp.~(\ref{element2})). 
Note that (\ref{rdDS1-f1}) 
(or, 
more precisely, its original form (\ref{dDS1-f1})) admits 
the simple vector generalization (\ref{dvDS1-f1}). 
It is shown that 
the two elementary flows (\ref{rdDS1-f1}) and (\ref{second_flow_cc}) 
(or, more generally, (\ref{dDS1-f1}) and (\ref{second_flow}))
commute under a natural  choice of the ``constants'' of integration. 
Thus, we can 
take a linear combination of them 
to obtain (\ref{semi-discrete_DS}), 
which provides 
an integrable semi-discretization 
of the Davey--Stewartson system (\ref{continuousDS}). 
By 
changing the independent variables as in (\ref{Galilean-like2}), 
we 
convert the 
semi-discrete elementary Davey--Stewartson 
flow 
(\ref{rdDS1-f1}) 
to the system (\ref{sd_YO1}), 
where the tilde of the continuous independent variables is omitted. 
The system (\ref{sd_YO1}) gives 
an integrable semi-discretization of 
the \mbox{$(2+1)$}-dimensional Yajima--Oikawa system (\ref{2DYO}),  
%
%
%
which is 
essentially equivalent to the semi-discrete system 
recently proposed by G.-F.~Yu and Z.-W.~Xu~\cite{Yu15}. 

We finally 
remark that 
another integrable discretization 
of the Davey--Stewartson system (\ref{continuousDS}) 
can be found in~\cite{TD11} (also see some preceding 
results in~\cite{Hu06,Hu07}), 
wherein both 
spatial variables $x$ and $y$ are discretized.


\end{document}